\documentstyle[jkas_eama6]{article}

\runningauthor{KO \& IM}
\runningtitle{COLOR GRADIENTS OF ELLIPTICAL
GALAXIES}
\beginpage{1}
\endpage{3}

\begin{document}

\title{OPTICAL-NEAR INFRARED COLOR GRADIENTS OF ELLIPTICAL GALAXIES AND THEIR ENVIRONMENTAL DEPENDENCE}

\author{JONGWAN KO AND MYUNGSHIN IM}

\address{Astronomy Program, School of Earth and Environmental Sciences, Seoul National University,
Seoul, 151-742, Korea \\
{\it E-mail: jwko@astro.snu.ac.kr}}

\abstract{
  We have studied the environmental effect on optical-NIR
 color gradients of 273 nearby elliptical galaxies. 
  Color gradient is a good tool to study the evolutionary
 history of elliptical galaxies, since the steepness of the color
 gradient reflects merging history of early types. When an elliptical
 galaxy goes through many merging events, the color gradient can 
 be get less steep or reversed due to mixing of stars.
  One simple way to measure color gradient is to compare half-light
 radii in different bands. We have compared the optical and near infrared
 half-light radii of 273 early-type galaxies from Pahre (1999).
 Not surprisingly, we find that
 r$_{e}$(V)s (half-light radii measured in V-band) are in general
 larger than r$_{e}$(K)s (half-light radii measured in K-band).
  However, when divided into different environments, we find that
 elliptical galaxies in the denser environment have gentler  
 color gradients than those in the less dense environment. 
  Our finding suggests that elliptical galaxies in the dense environment
 have undergone many merging events and the mixing of stars 
 through the merging have created the gentle color gradients.
}

\maketitle

\section{INTRODUCTION}
 

   Many studies have shown that stars in an elliptical galaxy gradually
 become redder toward the center, and this is called ``color gradient''
 (e.g., Franx \& Illingworth 1990; Peletier et al. 1990; Tamura \& Ohta 2003).
  It is known that the most dominant cause of the color gradient is a
 radial change of the stellar metallicity, i.e., metallicity gradient
 although there might be a small correction due to age gradient
 (e.g., Hinkley \& Im 2001). The color gradient becomes more 
 prominent when a long wavelength baseline is covered, such as
 the optical versus near-infrared bands (Hinkley \& Im 2001;
 La Barbera et al. 2003).

  Various galaxy formation models explain how the color gradient 
 (or equivalently the metallicity gradient) arises.
  Models suggest that elliptical galaxies form largely in two ways ---
 via monolithic collapse of the proto-galactic gas at high redshift 
 (``monolithic collapse model''; Eggen, Lynden-Bell, \& Sandage 1962;
 Larson 1975),
 or merging of gas-rich spiral galaxies at a more recent epoch 
 (``merging model''; Kauffmann,
 White, \& Guiderdoni 1993; Baugh, Cole, \& Frenk 1996). Either way,
 the steep gravitational well in the inner part of the elliptical galaxies
 allow the metal-enriched interstellar gas stay there for successive
 star formation events, which enriches the metallicity of 
 the surrounding gas even more. In contrast, supernovae explosions
 can easily blow away the surrounding ISM gas prohibiting 
 more stars to form in the outer part.
  Therefore, the metallicity gets enriched more in the inner part than 
 in the outer part, and this is considered to be the mechanism to produce
 the observed color-gradient in elliptical galaxies.
   Theoretical works suggest that the metallicity gradient
 can be as steep as $\Delta log(Z)/\Delta log(R) = -0.5$ to -1 under
 this kind of galaxy formation out of gas-rich systems (Larson 1974;
 Carlberg 1984). The observed steepness of the color gradients is 
 broadly consistent with  the theoretically computed values.
  If not many events happen after the initial formation of elliptical galaxies
 as the monolithic collapse model suggests, the color (or metallicity) gradient
 will stay at this initial state. However, under merging models,
 the situation changes. The merging activity complicates the color-gradient
 in many ways. When gas-poor elliptical galaxies merge, stars in the galaxies
 get mixed up, diluting the color-gradient. The collision between 
 a gas-poor elliptical galaxy and gas-rich satellite galaxy can lead to 
 sprinkling of young stars in the inner part of the elliptical galaxy,
 potentially creating a flat or a reversed color-gradient. There is already
 some observational evidences for the reversed color-gradient in elliptical
 galaxies (e.g., Im et al. 2001). However, it still remains as uncertain 
 how significantly merging activities play in the evolution and the formation
 of elliptical galaxies.

  Recently, hierarchical galaxy formation models put forward the 
 environment-dependent galaxy formation picture, where elliptical
 galaxies in high density environment form early and go through
 more merging events than elliptical galaxies in low density environment.
  Interestingly, spectral properties and colors of nearby ellipticals
 from the Sloan Digital Sky Survey (SDSS) turned out not to be 
 strongly dependent on environments (Bernardi et al. 2003).
 Such a finding, however,
 does not exclude a possibility that the present-day galaxies were
 assembled through merging of gas-poor galaxies or not - the hierarchical
 merging model is still viable. 
  As an attempt to test the merging history of nearby early-type
 galaxies, we have studied optical-NIR color gradients of nearby
 elliptical galaxies in different environments.

\section{DATA}

  Color gradient shows up as a change in half-light
 radii at different wavelengths.
  Half-light radii
 of early-type galaxies decrease systematically from a short wavelength to
 a long wavelength because of the color gradients.
   Furthermore, the difference in size can be substantial when the size in
 optical is compared with that in near infrared, i.e., when 
 a long wavelength baseline is explored.
 Therefore, the comparison of the optical vs near infrared sizes can
 be a simple and powerful way to measure color gradients
 in many elliptical galaxies
 (Sparks \& J{\o}rgensen 1993; Pahre et al. 1998). 

  In order to investigate the optical-near infrared color gradient, we
 have compared the $V$-band and $K$-band effective radii of 273
 elliptical galaxies from Pahre (1999).
   Pahre (1999) has compiled a list of 273 elliptical galaxies 
 is compiled for which optical structural parameters and the velocity
 dispersion measurements are available from literatures.
  For these ellipticals, they have obtained NIR $K$-band images, and
 derived $K$-band structural parameters.
 Using these galaxies, they have studied
 the fundamental plane relation in NIR.
  For our study, we have used the effective radii listed in 
 Pahre (1999) for $V$-band (hereafter, $r_{e}(V)$,
  and for $K$-band (hereafter, $r_{e}(K)$. We use the effective 
 radii measured using circular apertures.

\section{RESULTS}

\begin{figure}[h]
\centerline{\epsfysize=6cm\epsfbox{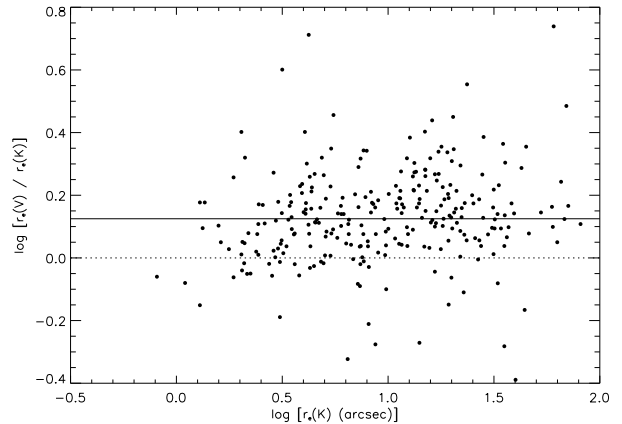}}
\caption{ The ratio of effective radius in V-band versus K-band of 273
galaxies from Pahre (1999). The solid line represents the median
value (0.125 dex).  }
\end{figure}

\begin{figure}[h]
\centerline{\epsfysize=6.5cm\epsfbox{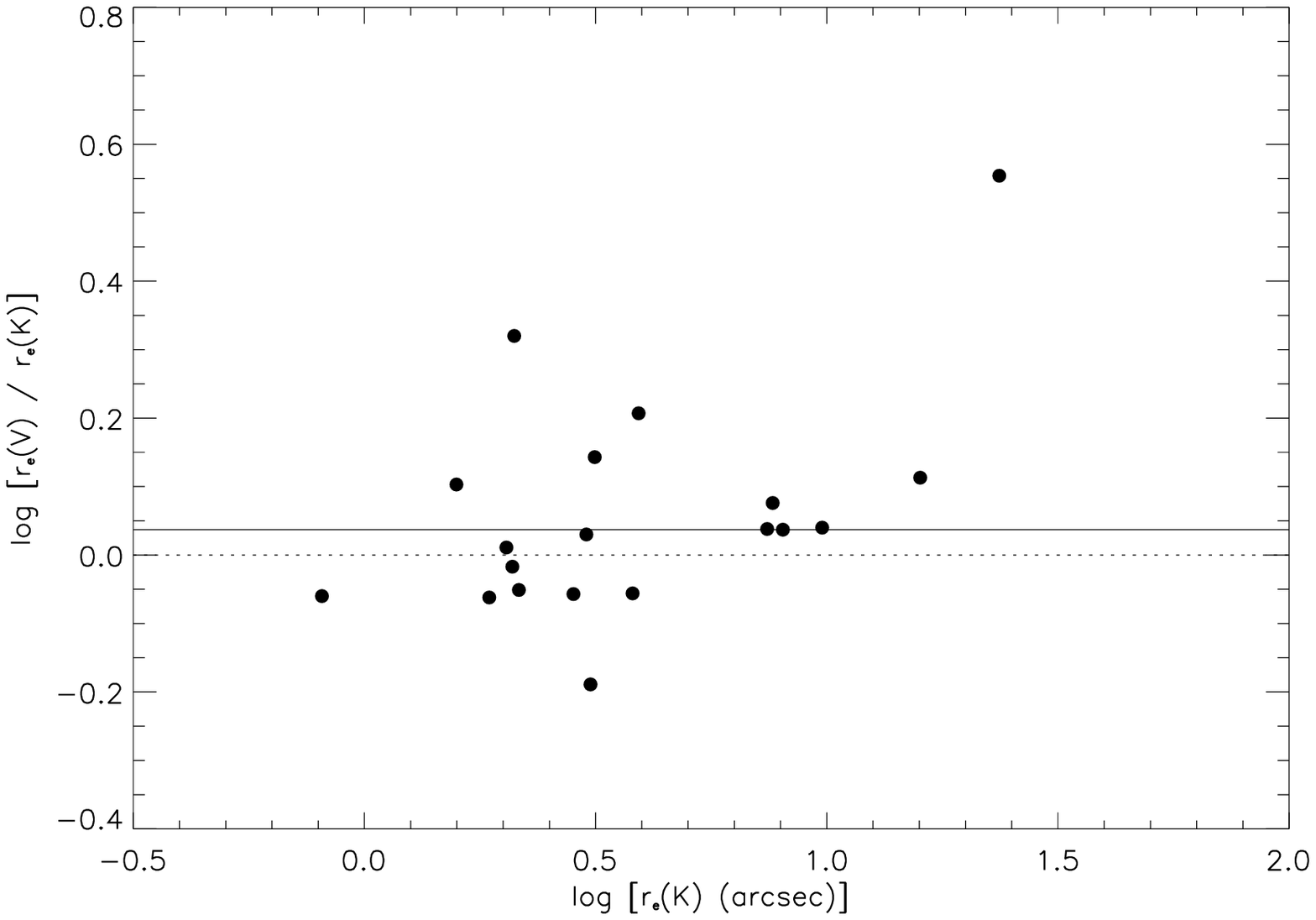}}
\centerline{\epsfysize=6.5cm\epsfbox{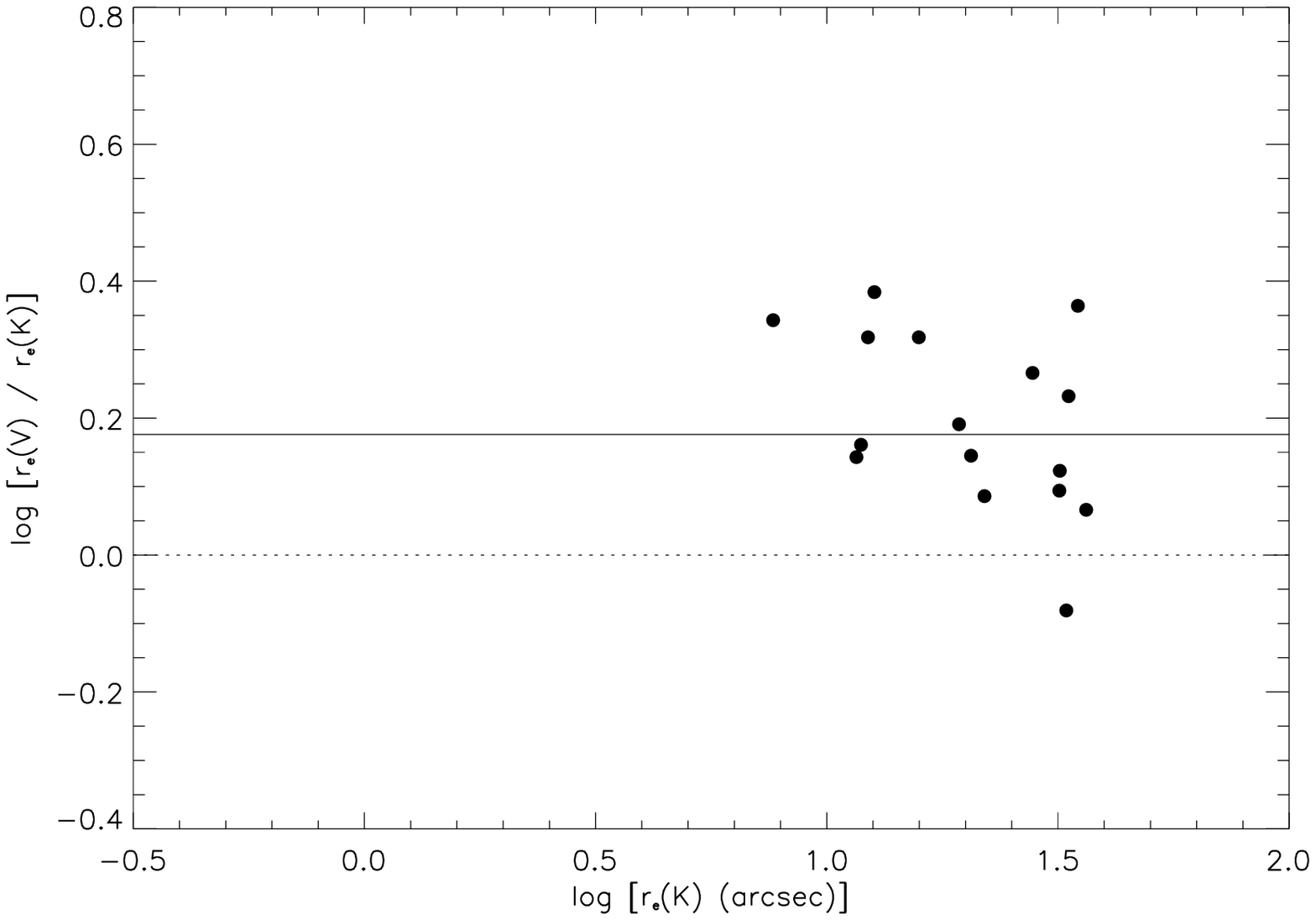}}
\centerline{\epsfysize=6.5cm\epsfbox{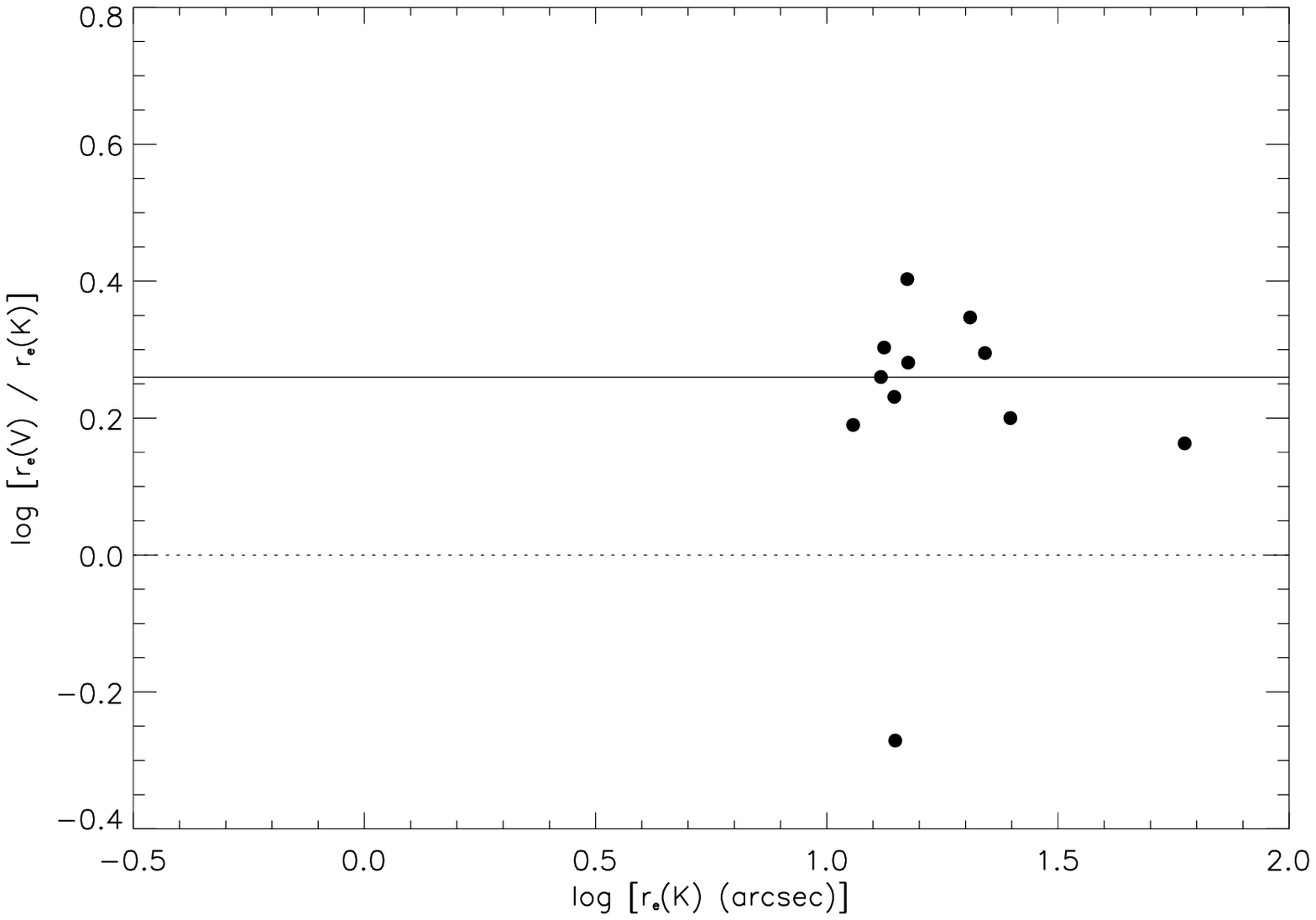}}
\caption{ Comparison of the effective radius ratio in different
environment.\emph{Top}, the rich cluster Abell 2199 (the median is
0.037 dex); \emph{middle}, the poor cluster Fornax (the median is
0.176); \emph{bottom}, field galaxies (the median is 0.260)}
\end{figure}

  First, we compare r$_{e}$(V) vs r$_{e}$(K) of 273 sample
 galaxies (Fig. 1). 
  The median of the ratio of r$_{e}$(V) and r$_{e}$(K) is
 1.33 , in other words, r$_{e}$(V) is about 33\% larger than
 r$_{e}$(K). Using the result of Sparks \& J{\o}rgensen (1993), the
 scale-length difference ($\Delta$s = $\Delta$r$_{e}$/r$_{e}$) can be
 converted into an isophotal color gradient of $\Delta(\mu_{V} -
 \mu_{K})=\beta\Delta\log$r, with $\Delta$s $\sim\beta$ = -0.33 mag
 arcsec$^{-2}$. 
  We find that some elliptical
 galaxies show a very large difference between optical effective radius 
 and near infrared effective radius. The difference can be more than
 a factor of 2.
 Some elliptical galaxies show positive color gradients, which
 means that stars are redder toward the outer part as merging 
 models suggest. 
  The optical-NIR color gradient from our study seems rather steep,
 compared to the expected size difference of 
 about 10-20 \% using the observed metallicity gradient 
 from the optical photometric and spectroscopic studies.
  There is a weak trend in Fig. 1 that galaxies with small 
 apparent sizes have smaller $r_{e}(V)/r_{e}(K)$ values. This is
 probably due to the difficulty of determining sizes when
 galaxy sizes are similar to the seeing size.

  Next, in Fig.~2, we compare color gradients of elliptical galaxies in
 different environments. Fig.~2 shows the ratio $r_{e}(V)/r_{e}(K)$ for 
 elliptical galaxies in a rich cluster Abell 2199 (the top panel),
 in a poor cluster, Fornax (the middle panel), and in field (the bottom panel).
  The median effective radii ratio for
 elliptical galaxies in the nearby rich cluster Abell 2199 is smaller
 than the same quantity in a poor cluster, Fornax. Also, we find that
 color gradients of cluster galaxies are much smaller than field
 galaxies. These results suggest that there is a relationship between
 the steepness of color gradients and environment: elliptical
 galaxies in the lower density regions tend to have steeper color
 gradient than ellipticals in the higher density regions. We have examined
 other elliptical galaxies in field or in other clusters, and 
 find a similar trend (Im \& Ko 2005, in preparation).

\section{DISCUSSION AND FUTURE WORK}

  The cause of the environmental dependence can be understood as
 the following. Hierarchical galaxy formation models predict that galaxy
 evolution proceeds differently depending on the environment.
  Early-types, when they formed first, are expected to have steep
 color gradients, since star formation activities are longer in the
 inner parts by abundant cold gas. Such prolonged star-forming
 activities could make stars in the inner parts more metal-rich,
 i.e., redder. After undergoing merging events, the color gradient of
 progenitor galaxy gets watered down as a consequence of mixing of
 stellar populations. Therefore, color gradients of cluster galaxies
 are gentler than field galaxies, since early-types in the cluster
 environment have gone through more merger events than in the
 field environment in such a model. 

  The rather large values for $r_{e}(V)/r_{e}(K)$ in Fig.~1 are 
 harder to understand. There is no question that the metallicity
 gradient is contributing here. We find that some galaxies with
 extremely large $r_{e}(V)/r_{e}(K)$ show the evidence for dust,
 therefore dust-extinction is another possible contributing factor
 (e.g., Wise \& Silva 1996).  
   We have examined another possibility that our result might be caused
 by a measurement error in the literature. To check for such a 
 possibility, we have used our own observation
 (imaging observation in optical bands using 1 m telescope at Lemmon and 1.5 m
 telescope at Maidanak) and archival data (2MASS), and derived 
 half light radii independently. Our independent 
 check finds no strong systematic bias in the published optical 
 sizes of galaxies in Pahre (1999), although we find that
 the literature values of $r_{e}$ for some ellipticals with 
 large $r_{e}(V)/r_{e}(K)$ turned 
 out to be wrong.  We plan to
 extend our analysis to a much larger sample using SDSS data set.
 A more complete analysis of this work will be presented elsewhere
 (Im \& Ko 2005, in preparation).

\acknowledgements

  This research was supported by the Brain Korea 21 (BK21)
 program, and the grant No.R01-2005-000-10610-0 from
 the Basic Research Program of the Korea Science \& Engineering Foundation.


\begin{references}

\reference{} Baugh, C. M., Cole, S., \& Frenk, C. S. 1996, MNRAS, 282, 27L

\reference{} Bernardi, M., et al. 2003, AAS, 202, 5104

\reference{} Carlberg, R. G. 1984, ApJ, 286, 416

\reference{} Eggen, O. J., Lynden-Bell, D., \& Sandage, A. R. 1962, ApJ, 136, 748

\reference{} Franx, M., \& Illingworth, G. 1990, ApJ, 359, L41

\reference{} Hinkley, S., \& Im, M. 2001, ApJL, 560, L41

\reference{} Im, M., et al. 2001, AJ, 122, 750

\reference{} Kauffmann, G., White, S. D. M, \& Guiderdoni, B. 1993, MNRAS, 264, 201

\reference{} La Barbera, et al. 2003, A\&A 409, 21

\reference{} Larson, R. B. 1975, MNRAS, 173, 671

\reference{} Larson, R. B. 1974, MNRAS, 166, 585

\reference{} Pahre, M. A. 1999, ApJS, 124, 127

\reference{} Pahre, M. A., de Carvalho, R. R., \& Djorgovski, S. G.
1998, AJ, 116, 1606

\reference{} Peletier, R. F., Davies, R. L., Illingworth, G. D.,
Davis, L. E., \& Cawson, M. 1990, AJ, 100, 1091

\reference{} Sparks, W. B., \& J{{\o}}rgensen, I. 1993, AJ, 105,
1753

\reference{} Tamura, N., Ohta, K. 2003, AJ, 126, 596

\reference{} Wise, M. W., \& Silva, D. R. 1996, ApJ, 461, 155

\end{references}
\end{document}